\documentclass[letterpaper]{article}
\usepackage[T1]{fontenc}
\usepackage[utf8]{inputenc}
\usepackage{lmodern}
\usepackage[margin=1in]{geometry}
\usepackage{setspace}
\usepackage[journal=jctcce,articletitle=true,maxauthors=0]{achemso}
\usepackage{authblk}
\usepackage{amsmath,amssymb}
\usepackage{graphicx}
\usepackage{array,booktabs,longtable,tabularx,makecell}
\usepackage{pdflscape}
\usepackage[table]{xcolor}
\usepackage{microtype}
\usepackage{newunicodechar}
\usepackage{caption}
\newunicodechar{₂}{\textsubscript{2}}
\usepackage{hyperref}
\mciteErrorOnUnknownfalse
\hypersetup{hidelinks,
  pdftitle={SATE: Transferable Spin-Adapted Neural Wavefunctions}}
\graphicspath{{figures/}}
\newcolumntype{L}[1]{>{\raggedright\arraybackslash}p{#1}}
\newcolumntype{C}[1]{>{\centering\arraybackslash}p{#1}}
\newcolumntype{Y}{>{\raggedright\arraybackslash}X}
\newcommand{\valerr}[2]{\makecell[c]{\(#1\)\\[-1pt]\(\pm #2\)}}

\title{SATE: Transferable Spin-Adapted Neural Wavefunctions}
\author[1,2]{Timothy Sereda}
\author[3]{Pallas Chou}
\author[1,*]{Morris Hsieh}
\author[1,*]{Emerson Hsieh}

\affil[1]{%
  Primateria\qquad
  \textsuperscript{2}University of South Dakota\qquad
  \textsuperscript{3}Stanford University
  \vspace{0.4em}%
}

\date{}

\begin{document}
\maketitle

\begingroup
\renewcommand{\thefootnote}{\fnsymbol{footnote}}
\footnotetext[1]{Correspondence to:
  \texttt{\{morris,emerson\}@primateria.com}}
\endgroup

\vspace{-3em}

\begin{abstract}

Accurate prediction of molecular reactivity requires both accurate and computationally efficient
description of electron correlation and spin across various molecular geometries and electronic
states. We introduce SATE, a \textbf{s}pin-\textbf{a}dapted \textbf{t}ransferable neural wavefunction architecture with
\textbf{e}ffective core potentials.
SATE combines a transferable neural wavefunction,
the spin-adapted antisymmetrization method, and effective core potentials into a single foundation
model architecture pretrained on compositions containing selected elements from hydrogen to bromine. After pretraining and adaptation, our model is able to predict 3d transition-metal spin-state gaps, bond dissociation energies, ionization energies, and a reaction energy with absolute deviations of 2.1--8.7 mHa from reference values, based on two-seed means. Effective core potentials reduce the number of explicitly treated electrons, while spin-adapted antisymmetrization enforces the prescribed total spin. By testing our model with 3d transition metal chemistry, we evaluate our model's description of strong electron correlation and competing electronic states in such systems. We find that pretraining improves most tested component energies at matched adaptation budgets, while improvements in spin-state gaps, dissociation energies, ionization energies and the reaction energy remain system dependent. Ultimately, our work lays the foundation for more accurate and transferable modeling of molecular systems by combining the computational efficiency of effective core potentials with explicit control over total spin.

\end{abstract}

\section{Introduction}

Accurate predictions of molecular reactivity require electronic structure methods that remain
reliable as bonds break and competing electronic states change in energy. Transition-metal compounds present a distinct computational challenge. Because their partially occupied d shells often generate near-degenerate electronic configurations and multiple low-lying spin states, predicting accurate relative energies requires a method that simultaneously accounts for both static and dynamic electron correlation.~\cite{sharma2014}. The accuracy of density functional theory (DFT) and
single-reference coupled-cluster, such as coupled cluster with singles, doubles, and
perturbative triples (CCSD(T)), depends on the system and electronic state. For example, benchmark studies on DFT and CCSD(T) show substantial method dependence in transition-metal spin-state energetics~\cite{sse2024}.
Multireference approaches can describe this configurational complexity but often require
substantial computational effort.~\cite{sharma2014} More tractable multireference methods would have broad implications for modeling catalysis~\cite{vitillo2022multireference}, energy conversion~\cite{zimmerman2010singlet}, and biological systems.~\cite{sharma2014}

The versatility of deep neural networks provides a route to approximate
correlated many-electron wavefunctions directly, including those that are strongly correlated. In
deep quantum Monte Carlo (deep QMC), a neural network wavefunction is optimized through variational
Monte Carlo, allowing electron correlation to be described without relying on an approximate
exchange-correlation functional or a truncated coupled-cluster expansion. For example, FermiNet
\cite{ref1} and Psiformer \cite{ref2} construct determinants from learned orbitals that depend on
the full electronic configuration, while PauliNet augments a multireference Hartree-Fock baseline
with neural-network Jastrow factors and backflow transformations~\cite{pauli2020}. Obtaining accurate wavefunctions
nevertheless requires us to optimize wavefunction parameters separately for each new target,
motivating methods that reuse learned electronic structures across related systems.

Transferable neural wavefunctions seek to reduce this repeated computational effort by sharing
learned representations across molecular systems. A model pretrained across multiple geometries and
compositions can provide initialization for subsequent variational optimization, which potentially
reduces the compute required to obtain an accurate wavefunction for a new target. Early work
demonstrated parameter sharing across nuclear geometries~\cite{scherbela2022}. Subsequent work introduced a transferable mapping from Hartree-Fock orbitals to correlated neural orbitals, demonstrating that pretraining on
smaller molecules could accelerate calculations on larger compounds~\cite{scherbela2024}. Orbformer subsequently scaled chemical pretraining to 22,350 molecular structures and demonstrated adaptation to bond dissociation
and reaction processes~\cite{ref3}.

Extending this framework to transition-metal chemistry requires broader chemical coverage and
explicit control of total spin. Orbformer's published pretraining curriculum, named the Light Atom Curriculum (LAC), comprises H, Li, B, C, N, O, F \cite{ref3}. In transition-metal compounds, the relative energies and electronic character of low-lying spin states can change during bond formation and dissociation. Comparing these states requires wavefunctions with the intended total spin. Specifying the numbers of spin-up and spin-down electrons fixes the spin projection (\(M_S\)), but does not generally enforce the total spin (\(S\)). Consequently, energy minimization can favor a lower-energy state with unintended spin character, compromising the calculation of spin-state gaps and the interpretation of reaction pathways. The spin-adapted antisymmetrization method (SAAM) addresses this problem by jointly antisymmetrizing a neural spatial function and a prescribed spin eigenfunction, preserving the target total spin by construction \cite{ref4}.

Extending chemical coverage to heavier atoms also increases the computational burden of treating
electrons explicitly. Effective core potentials (ECPs) provide an established route to reducing the
computational cost of deep QMC by replacing selected core electrons with effective electron-ion
interactions~\cite{ref5,liECP2022}. Semilocal effective core potentials have been incorporated into FermiNet~\cite{liECP2022}, while local pseudo-Hamiltonians can further reduce computational cost by avoiding nonlocal integration, enabling calculations on iron--sulfur clusters~\cite{fuPH2026}. These developments motivate
incorporating ECPs into a chemically pretrained, explicitly spin-adapted architecture.

Here, we introduce SATE, a \textbf{s}pin-\textbf{a}dapted \textbf{t}ransferable neural wavefunction architecture with \textbf{e}ffective
core potentials. SATE combines Orbformer's spatial representation, SAAM-based spin adaptation, and effective-core Hamiltonians in a common neural network architecture. We pretrain the model
across an expanded chemical curriculum and investigate whether it provides useful initializations
for spin-adapted calculations on compositions excluded from pretraining. We then use
molecular-fragment energy differences to test the resulting wavefunctions with chemical observables. By combining broader chemical coverage with explicit control over total spin, SATE provides a framework for comparing molecular energies across geometries, compositions, and spin states.

\section{Results}\subsection{SATE: an architecture for transferable spin-adapted wavefunctions}

\begin{figure}[htbp]
  \centering
  \includegraphics[width=0.50\linewidth]{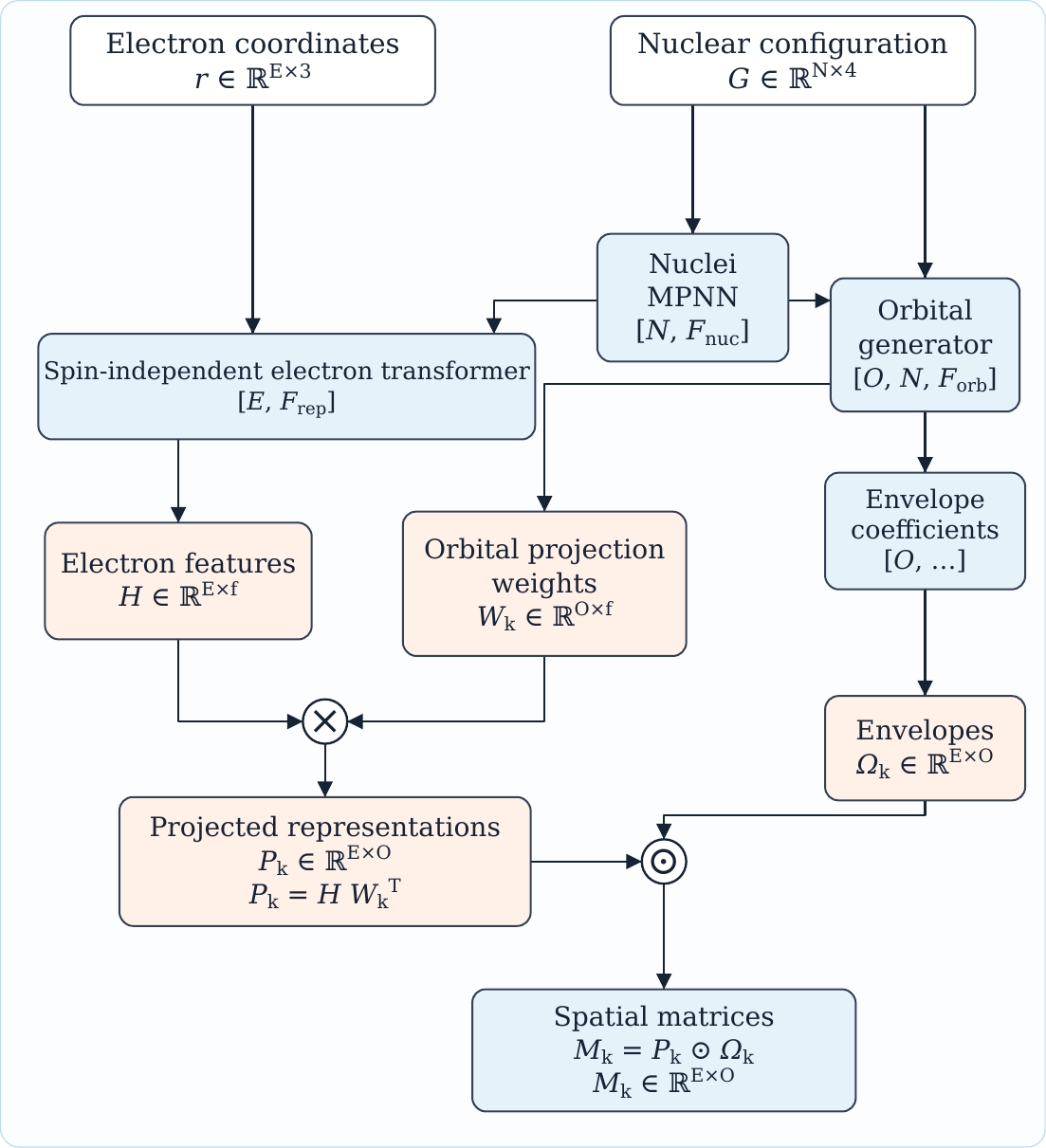}
  \caption{\textbf{SATE Architecture.}
    We use an Orbformer-derived permutation-equivariant spatial network but do not distinguish electrons by
their spin labels. Exchanging two electron positions exchanges the corresponding rows of the spatial matrix. This allows us to introduce spin through prescribed spin functions and jointly
antisymmetrize the spatial and spin contributions.
  }
  \label{fig:spatial-architecture}
\end{figure}

Our wavefunction combines Orbformer's nuclear-conditioned spatial representation \cite{ref3} with
spin-adapted antisymmetrization \cite{ref4} and correlation-consistent effective core potentials~\cite{ref5}. A
nuclear message-passing network constructs features from nuclear species and geometry. The features
condition both the electron transformer and the orbital generator. The transformer produces a
feature vector for each electron that depends on the surrounding nuclei and the positions of all
electrons (Figure~\ref{fig:spatial-architecture}).

Transferability comes from using shared neural-network parameters to generate orbitals conditioned
on the nuclear configuration. Multi-system pretraining teaches the network representations that can
be used as a starting point for new molecules.

\begin{figure}[htbp]
  \centering
  \includegraphics[width=0.50\linewidth]{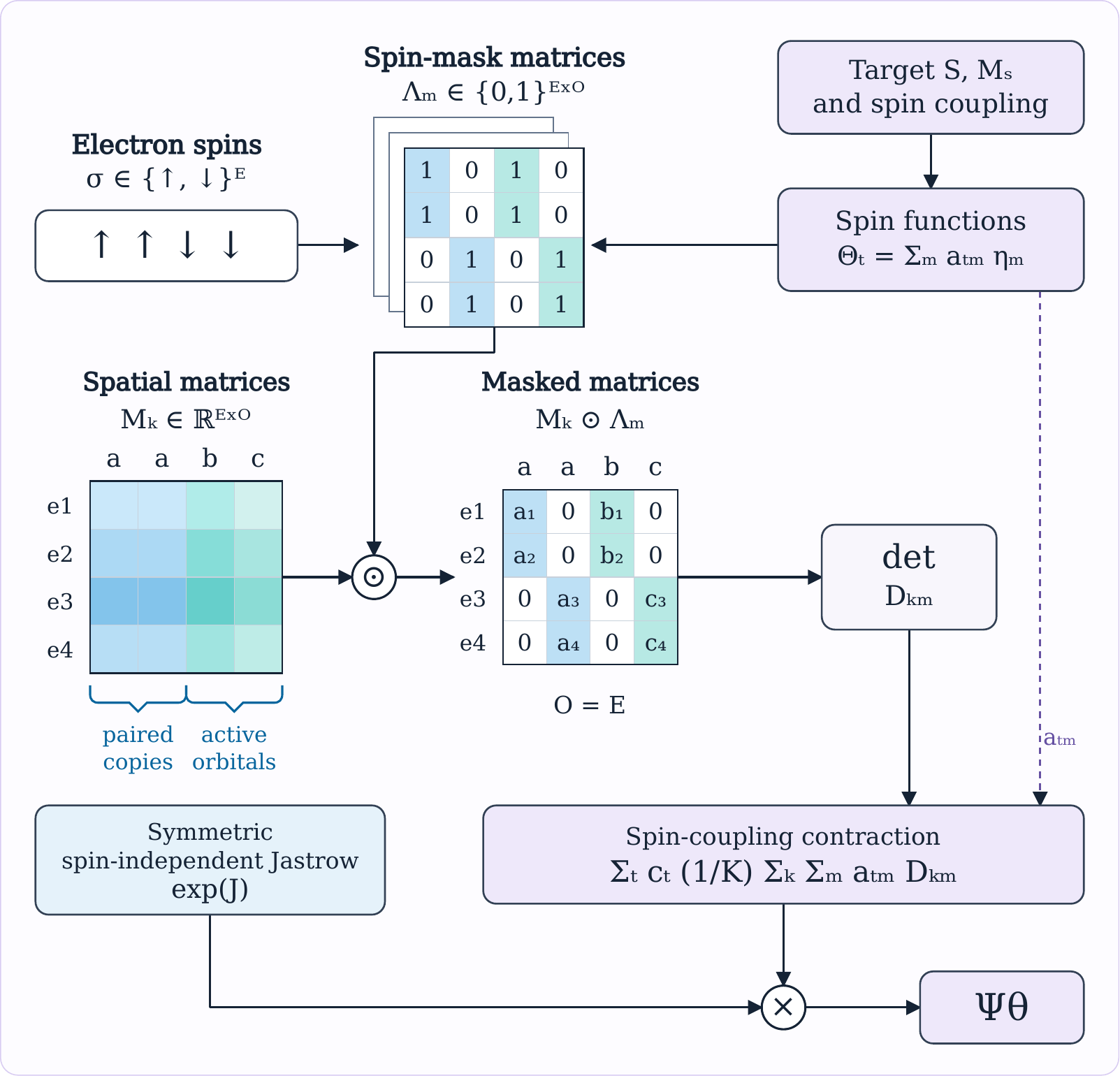}
  \caption{
    \textbf{Spin-adapted wavefunction construction.}
    \(G_\theta^{(k)}\) is a product of generalized spatial orbitals from
    component \(k\), \(\Theta_t^{S,M_S}\) is a prescribed spin-coupling
    function, and \(\hat{\mathcal A}\) is the joint antisymmetrizer.
    The learned coefficients \(c_t\) mix retained spin-coupling functions
    with the same target spin. They are normalized and shared across
    all \(K\) spatial components.
  }
  \label{fig:spin-adaptation}
\end{figure}

Specifying the number of up- and down-spin electrons fixes the spin projection \(M_S\), but generally does not
 fix the total spin \(S\). Following SAAM~\cite{ref4}, we construct the wavefunction by combining spatial
functions with prescribed spin eigenfunctions and antisymmetrizing jointly in space and spin (Figure~\ref{fig:spin-adaptation}):
\[
\Psi_\theta^{S,M_S}
=e^{J_\theta}\sum_{k,t}\frac{c_t}{K}\,
\hat{\mathcal A}
\left[G_\theta^{(k)}\Theta_t^{S,M_S}\right].
\]

Each prescribed function is a fixed linear combination of up/down assignments. We implement each
assignment as a mask on the neural spatial matrix and evaluate the resulting determinant. Combining
these determinants with the prescribed coupling coefficients produces a spin-adapted wavefunction.
When multiple spin-coupling functions with the same target \(S\) and \(M_S\) are retained, their relative
weights are learned during optimization.

\subsection{Evaluation of transferability and spin control}

We investigate whether pretraining across a range of chemical systems provides a transferable
initialization for spin-adapted neural wavefunctions on new molecular compositions. Pretrained and
cold-start calculations assess the optimization required to obtain low-energy wavefunctions on
molecular compositions excluded from pretraining. Molecular-fragment energy
differences attempt to provide chemical validation
of the resulting wavefunctions. Independent evaluations of frozen checkpoints assess reliability of
reported energies.

\subsection{Spin adaptation enforces the prescribed total spin}

We first assess total-spin control. For O₂ and Fe\textsuperscript{+}, the spin-adapted wavefunctions retain the prescribed values of \(\langle\hat S^2\rangle\), while the corresponding fixed-\(M_S\) Orbformer-based control calculations exhibit spin contamination, particularly for the lower-spin targets (Table~\ref{tab:spin-control}). These measurements verify the spin construction independently of the accuracy of the calculated energies.
\par\medskip
\begin{table}[htbp]
\centering
\caption{Control of total spin.}
\label{tab:spin-control}
\small\setlength{\tabcolsep}{5pt}\renewcommand{\arraystretch}{1.3}
\begin{tabularx}{\linewidth}{@{}lYYYC{3.3cm}C{1.9cm}@{}}
\toprule
\textbf{System} & \textbf{SATE prescribed $S$} & \textbf{Fixed $M_S$ (both)} & \textbf{SATE $S(S+1)$} & \textbf{Orbformer-based $\langle S^2\rangle$} & \textbf{SATE $\langle S^2\rangle$}\\
\midrule
O$_2$ & 0 (singlet) & 0 & 0 & $1.0245\pm0.0047$ & $0^{\dagger}$\\
O$_2$ & 1 (triplet) & 1 & 2 & $2.0100\pm0.0008$ & $2^{\dagger}$\\
Fe$^+$ & $3/2$ (quartet) & $3/2$ & 3.75 & $4.6918\pm0.0085$ & $3.75^{\dagger}$\\
Fe$^+$ & $5/2$ (sextet) & $5/2$ & 8.75 & $8.7563\pm0.0031$ & $8.75^{\dagger}$\\
\bottomrule
\end{tabularx}
\par\smallskip
$\dagger$ Rounded frozen 10k measurements. Maximum mean deviation from $S(S+1)$: \(<3\times10^{-12}\).
Total-spin measurements use frozen checkpoints after 10,000 optimization updates. Orbformer fixes \(M_S\), while SATE prescribes \(S\). Each state is evaluated using two separately seeded chains, each with 1,024 walkers, 1,200 burn-in sweeps, and 512 measurement records separated by 60 sweeps. FP64 pairing is performed on every 16th record. Initialization and target settings are given in \hyperref[app:evaluation-targets]{Appendix~5}.

\end{table}

\enlargethispage{2\baselineskip}
\subsection{Pretraining improves component energies at matched adaptation budgets}

We compare frozen energies after 20,000 accepted VMC adaptation updates using the same architecture for all initializations. Warm starts use the 100,000-update pretrained checkpoint; HF starts include 2,000 orbital-fitting updates before VMC.

Warm initialization yields lower two-seed mean central energies than random initialization for 20 of 21 distinct components and lower estimates than HF initialization for 13 of 20 components (Figure~\ref{fig:component-transfer}). The HF comparison excludes the O$_2$ triplet estimate with substantial drift during frozen evaluations. Several differences are comparable to their combined uncertainties. Target geometries and active-electron counts are provided in \hyperref[app:evaluation-targets]{Appendix~5}.

\subsection{Improvements in energy differences depend on error cancellation}
Improvements in component energies do not consistently result in more accurate observables (Figure~\ref{fig:observable-accuracy}; Table~\ref{tab:transfer-results}). For CuH, warm initialization lowers the molecular and Cu-atom energies relative to random  initialization by $71.61\pm62.20$ mHa and $31.59\pm22.13$ mHa, respectively. These unequal gains produce $D_e=108.70\pm3.85$ mHa, compared with $68.68\pm83.65$ mHa from random initialization and a reference of $103.11\pm1.14$ mHa. For CrH, warm initialization brings the dissociation energy from $162.06\pm10.40$ to $83.58\pm4.33$ mHa, close to the reference of $81.51\pm2.67$ mHa. Pretraining does not improve every observable: the Sc ionization energy has an absolute error of 8.29 mHa, compared with 1.63 mHa from HF initialization.

The corresponding Cu/H$_2$ reaction energy is $65.06\pm3.86$ mHa, compared with the reference of $71.39\pm1.14$ mHa. This reaction and the CuH binding energy share molecular and atomic components, so their errors are correlated. The warm Fe$^+$ quartet-sextet gap is $12.02\pm5.16$ mHa against a reference of $9.11$ mHa, while the Cr ionization energy is $244.24\pm2.90$ mHa against $248.66$ mHa.
\clearpage

\noindent
\begin{minipage}{\textwidth}
  \centering
  \includegraphics[
    width=\linewidth,
    height=0.75\textheight,
    keepaspectratio
  ]{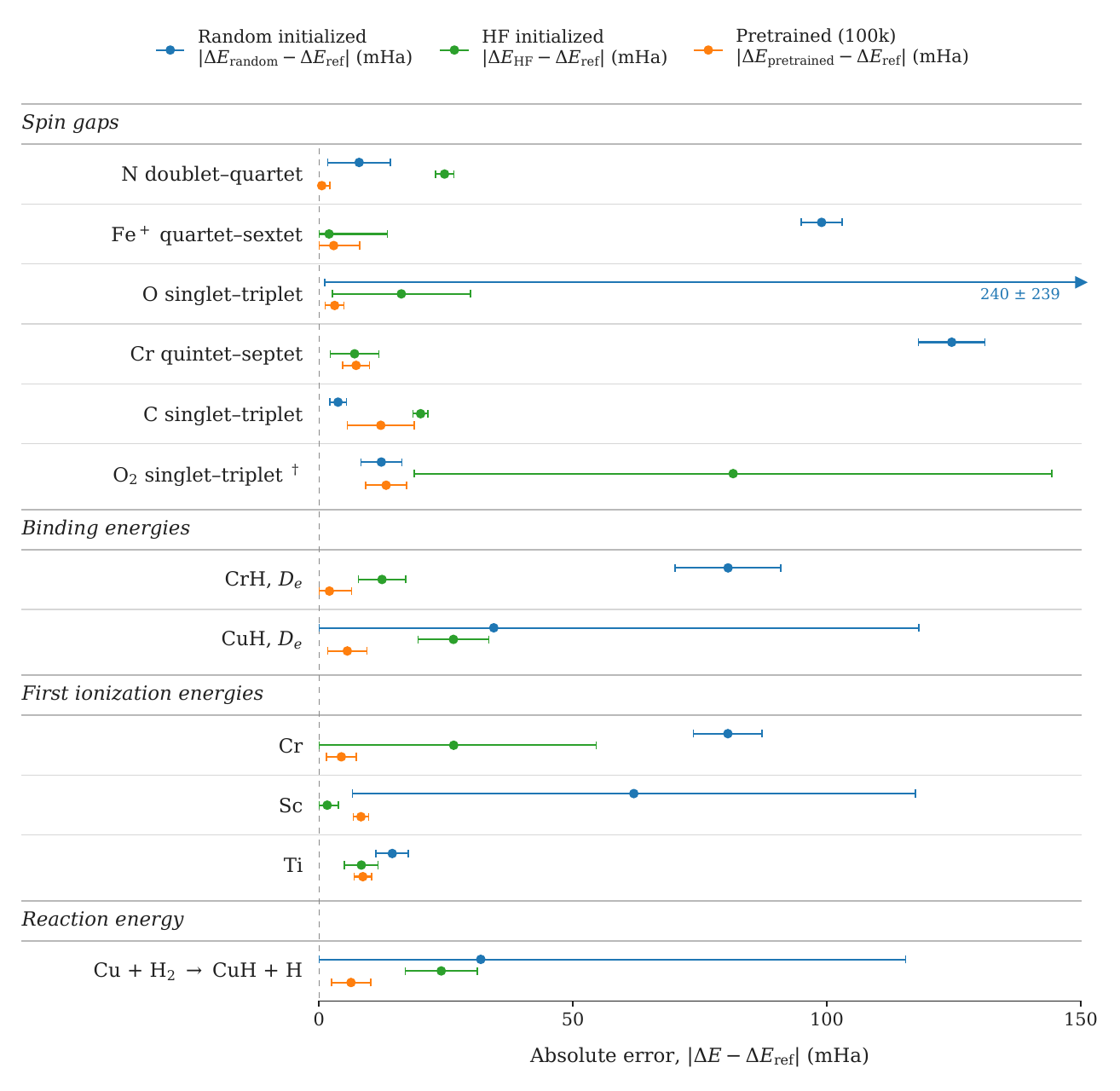}

  \captionof{figure}{\textbf{Observable accuracy.} Frozen observable errors after 20,000 adaptation updates, comparing random (blue), HF (green), and pretrained (100k; orange) initialization. Points show the absolute difference between the two-seed mean and the reference; observables are calculated within each seed before averaging. Uncertainties use the larger of the propagated Monte Carlo and between-seed standard errors. Reference uncertainties are not included. Binding and reaction energies exclude zero-point corrections. Arrows indicate intervals extending beyond 150 mHa. $\dagger$ The HF-initialized O$_2$ gap includes a triplet read with substantial energy drift during frozen measurement.
  }
  \label{fig:observable-accuracy}
\end{minipage}

\clearpage

\noindent
\begin{minipage}{\textwidth}
  \centering
  \includegraphics[
    width=\linewidth,
    height=0.75\textheight,
    keepaspectratio
  ]{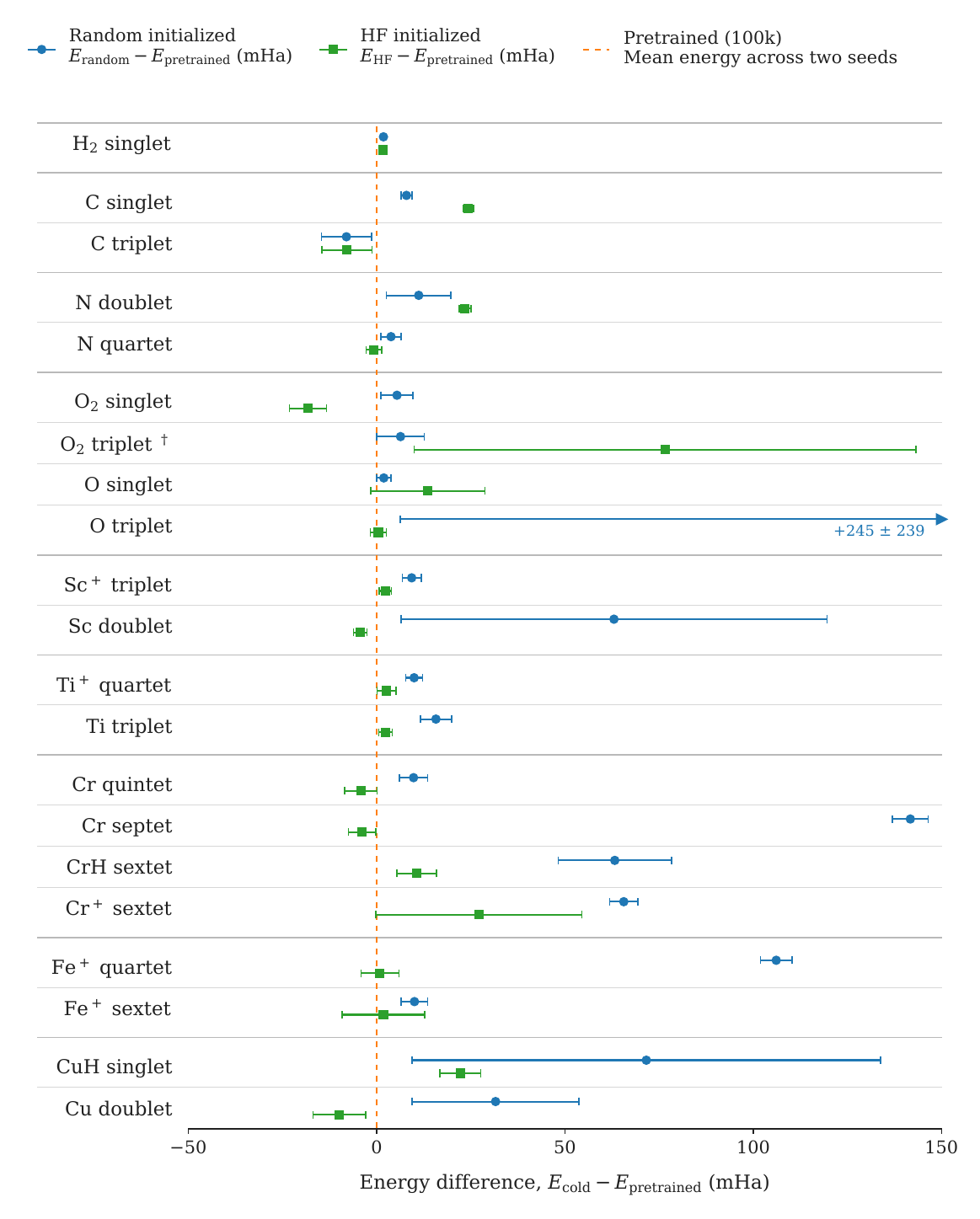}

  \captionof{figure}{\textbf{Component-energy improvement from pretraining.} 
  Frozen component-energy differences after 20,000 adaptation updates, comparing random (blue) and HF (green) initialization with the 100k pretrained initialization. Positive values indicate higher cold-start energies. Points compare two-seed means. Each bar includes uncertainty from both the cold and pretrained estimates with \(\sigma_{\Delta E} = \sqrt{\sigma_{\mathrm{cold}}^{2} + \sigma_{\mathrm{pretrained}}^{2}}\). Each mean uses the larger of its propagated Monte Carlo and between-seed standard errors. The right arrow marks an uncertainty interval extending beyond +150 mHa. $\dagger$ The HF-initialized O$_2$ triplet includes a read with substantial energy drift during frozen measurement.}
  \label{fig:component-transfer}
\end{minipage}

\clearpage
\section{Discussion}

Our results show that chemical pretraining can provide useful initializations for spin-adapted calculations on compositions excluded from the training curriculum, including transition-metal systems. At matched adaptation budgets, pretraining yields lower central energy estimates for most of the tested states and fragments, with a more consistent advantage over random initialization than over Hartree–Fock initialization. In its current form, SATE provides a common pretrained starting point for fixed-geometry calculations on atoms, ions, and small molecules at prescribed total spin. Using this approach, we computed spin-state gaps, ionization energies, bond dissociation energies, and the energy of a model reaction. SATE extends transferable wavefunction frameworks to include 3d transition-metal systems while enforcing the prescribed total spin.

In the tested O$_2$ and Fe$^+$ states, SATE retains the prescribed total spin, including for lower-spin targets where fixed-$M_S$ calculations exhibit spin contamination. 

A key next step is to quantify how initialization, model capacity and optimization strategy affect the computational cost of reaching a target accuracy. With reliable accuracy and transfer, a shared pretrained representation can reduce the repeated optimization needed to investigate bond activation, reaction energetics, and competing electronic states. Such calculations could support systematic comparisons of how substitution, charge, and metal ligation alter bonding and reactivity in organic molecules, catalysts, and enzyme active-site models. Studies of organic diradicals \cite{cramer1998bergman}, iron-oxo catalysis \cite{rice2024}, and cytochrome P450 \cite{yosca2013} illustrate how these chemical changes affect spin-state energetics and bond activation. Beyond molecular applications, SATE could be adapted to periodic boundary conditions, building on prior work on transferable neural wavefunctions for solids  \cite{gerard2025}. Furthermore, combining parameter reuse with total-spin control could support comparisons of competing spin states at semiconductor defects \cite{gordon2015}.

SATE is designed as a foundation model for spin-resolved electronic structure across chemical space. By combining broader chemical coverage with explicit control over total spin, SATE provides a framework for comparing molecular energies across geometries, compositions, and spin states.

\section{Methods}

\subsection{Transferable spatial representation}

The orbital generator supplies feature-to-orbital projection weights and electron-nucleus envelope
coefficients. Envelope decay exponents are trainable element-specific parameters shared across orbitals, spatial components, and geometries. For spatial component k, the matrix is:

\[
M_\theta^{(k)} = \left[H_\theta\left(W_\theta^{(k)}\right)^{\mathsf T}\right]
\odot\Omega_\theta^{(k)}.
\]
where \(H_\theta\) contains the electron features, \(W_\theta^{(k)}\) maps the features to orbital values,
and \(\Omega_\theta^{(k)}\) are the envelopes. The \(\odot\) symbol is element-wise multiplication.
Rows correspond to electrons and columns to spatial orbitals. Since the electron features depend on the
complete electron configuration, these are generalized orbitals rather than independent-particle orbitals.

For each doubly occupied spatial orbital, two columns share projection and envelope parameters. The
subsequent spin construction associates these copies with opposite spins, allowing double occupation
of that orbital. Although the columns are identical in the spatial matrix, their spin factors
distinguish them during antisymmetrization. Active orbital columns retain independent parameters
rather than being constrained to identical pairs.

\subsection{Spin-adapted wavefunction}

Each spin primitive is a product of one-electron spin functions and is represented computationally by a
spin-mask matrix. Fixed linear combinations of these primitives form independent spin-adapted coupling
functions with prescribed \(S\) and \(M_S\). Learned coefficients mix the retained coupling functions,
while \(K\) controls the number of spatial components.
\[
N_{\mathrm{primitive}}=\binom{n_{\mathrm a}}{n_{\mathrm a}/2-S},
\qquad
N_{\mathrm{coupling}}
=\binom{n_{\mathrm a}}{n_{\mathrm a}/2-S}
-\binom{n_{\mathrm a}}{n_{\mathrm a}/2-S-1}.
\]
Here, \(n_a\) is the number of active electrons, assuming \(M_S=S\). These functions are used to
generate the spin-tables for adaptation and SATE pretraining, counting the complete basis for the
chosen active space~\cite{ref4}. The pretraining curriculum (Table~\ref{tab:replacement-curriculum}) and transfer benchmarks use spin-coupling bases containing zero to seven active electrons, with the remaining electrons assigned to paired orbitals. Active-electron spins are coupled to the prescribed total spin.

The complete determinant sum has the prescribed total spin, while an individual determinant need not.
We evaluate the sum using logarithmic arithmetic and verify spin character by evaluating
\(\langle\hat S^2\rangle\). A symmetric, spin-independent Jastrow factor adds correlation without changing total spin.

A complementary approach, spin-adapted neural-network backflow (SA-NNBF), enforces total spin in a second-quantized representation~\cite{li2026sannbf}.

\subsection{Hamiltonians and variational optimization}

We work within the Born-Oppenheimer approximation, fixing the nuclei as point charges. Core electrons are approximated with an effective core potential and the remaining electrons are treated explicitly. In atomic units, the following equation describes the Hamiltonian of an atomic or molecular system:

\[
\hat H=-\frac12\sum_i\nabla_i^2+\sum_{i<j}\frac1{r_{ij}}
+\sum_{iA}\hat V_A^{\mathrm{ECP}}+E_{\mathrm{ion}}.
\]
where \(r_{ij}\) is the electron-electron separation, \(\hat V_A^{\mathrm{ECP}}\) contains the local and
nonlocal electron-ion interactions at center \(A\), and \(E_{\mathrm{ion}}\) is the repulsion between nuclei
with their effective charges. The nonlocal terms distinguish angular-momentum channels around each
nucleus. For the 3d transition metals, the Ne-core correlation-consistent effective core potentials (ccECPs) replace ten core electrons while retaining the
\(3s\) and \(3p\) semicore electrons explicitly, together with the \(3d\) and \(4s\) valence electrons~\cite{ref5}.
Spin-orbit coupling is omitted, so the Hamiltonian commutes with the total-spin operator \(\hat S^2\).
Molecular and fragment energies are calculated using the same effective-core Hamiltonian. During pretraining, hydrogen was treated all-electron. For the other curriculum systems, ccECPs were used~\cite{ref5,bennett2017ccecp,bennett2018ccecp,wang2019ccecp}. In the evaluation panel, O$_2$, C, N, O, and H$_2$ were treated all-electron, while transition-metal centers used ten-electron-core ccECPs.

For a wavefunction \(\Psi_\theta\) with parameters \(\theta\), variational Monte Carlo estimates the energy expectation
\[
E_\theta=\frac{\langle\Psi_\theta|\hat H|\Psi_\theta\rangle}
{\langle\Psi_\theta|\Psi_\theta\rangle}=E_{\pi_\theta}[E_L]
\]
where,
\[
\pi_\theta(r)=\frac{|\Psi_\theta(r)|^2}{Z_\theta},\qquad
E_L(r)=\frac{(\hat H\Psi_\theta)(r)}{\Psi_\theta(r)}.
\]
Here \(r\) denotes the complete set of electron coordinates and \(Z_\theta\) is the normalization integral.

Markov-chain electron position sampling of configurations estimates the energy and its parameter
gradients. During optimization, changes to \(\theta\) also change the target distribution \(\pi_\theta\).
We optimize the wavefunction parameters using Kronecker-factored approximate curvature (KFAC) \cite{ref8}.
Spatial derivatives required for the kinetic energy are evaluated using forward Laplacian automatic
differentiation~\cite{ref9} (see \hyperref[app:appendix-1]{Appendix~1}).

\subsection{Pretraining and adaptation protocol}
\begin{table}[htbp]
\centering
\caption{Model architectures and training protocols.}
\label{tab:model-comparison}
\footnotesize\setlength{\tabcolsep}{5pt}\renewcommand{\arraystretch}{1.12}
\begin{tabularx}{\linewidth}{@{}L{3.35cm}YYY@{}}
\toprule
& \textbf{SAAM}~\cite{ref4} & \textbf{Orbformer (LAC)}~\cite{ref3} & \textbf{SATE}\\
\midrule
Spatial backbone & LapNet~\cite{ref9} & Orbformer & Orbformer-derived\\
Layers $\times$ hidden width & $4\times256$ & $4\times256$ & $4\times256$\\
Spatial components, $K$ & 16 & 16 & 16\\
Prescribed total spin & Yes & No & Yes\\
Multi-system pretraining curriculum & None (Intrasystem) & 22,350 light-element configurations & 30,270 training entries; 137 compositions\\
Training budget & 5k or 20k HF initialization updates; 100k-200k VMC updates per stage (2-3 stages) & 800k pretraining updates across three stages & 100k pretraining updates; 20k adaptation updates per seed\\
Walker pruning & No & Yes & No\\
Sampling & Random-walk Metropolis & ULA $\to$ MALA; then MALA between updates & Random-walk Metropolis 30 moves between updates, walker refreshes every 40 updates and 1,000 equilibration sweeps.\\
Sampling sweeps between optimization updates & 30 & 60 & 30\\
\bottomrule
\end{tabularx}
\end{table}
\par\medskip
\begin{table}[htbp]
\centering
\footnotesize
\setlength{\tabcolsep}{4pt}
\renewcommand{\arraystretch}{1.12}
\caption{Pretraining curriculum.}
\label{tab:replacement-curriculum}
\begin{tabularx}{\linewidth}{@{}>{\raggedright\arraybackslash}p{2.3cm}rrr>{\raggedright\arraybackslash}X@{}}
\toprule
Subset & Composition \# & Geometries & Spin configurations & Compositions \\
\midrule
Light elements & 65 & 5,000 & 5,270 & \(\mathrm{B}\), \(\mathrm{B_{2}CH_{2}}\), \(\mathrm{BCH_{2}F}\), \(\mathrm{BCH_{3}}\), \(\mathrm{BCH_{3}O}\), \(\mathrm{BCH_{4}N}\), \(\mathrm{BF_{3}}\), \(\mathrm{BH_{2}F}\), \(\mathrm{BH_{3}}\), \(\mathrm{BH_{3}O_{2}}\), \(\mathrm{Be}\), \(\mathrm{C_{2}}\), \(\mathrm{C_{2}H_{2}}\), \(\mathrm{C_{2}H_{3}F}\), \(\mathrm{C_{2}H_{3}Li}\), \(\mathrm{C_{2}H_{3}N}\), \(\mathrm{C_{2}H_{4}}\), \(\mathrm{C_{2}H_{4}O}\), \(\mathrm{C_{2}H_{5}Li}\), \(\mathrm{C_{2}H_{6}}\), \(\mathrm{C_{2}Li_{2}}\), \(\mathrm{CFN}\), \(\mathrm{CH_{2}FLi}\), \(\mathrm{CH_{2}O}\), \(\mathrm{CH_{2}O_{2}}\), \(\mathrm{CH_{3}F}\), \(\mathrm{CH_{3}Li}\), \(\mathrm{CH_{3}N}\), \(\mathrm{CH_{3}NO}\), \(\mathrm{CH_{4}LiN}\), \(\mathrm{CH_{4}O}\), \(\mathrm{CH_{5}N}\), \(\mathrm{CHF}\), \(\mathrm{CHFO}\), \(\mathrm{CHLiO}\), \(\mathrm{CHN}\), \(\mathrm{CLiN}\), \(\mathrm{CO}\), \(\mathrm{F}\), \(\mathrm{F_{2}}\), \(\mathrm{H_{2}F_{2}}\), \(\mathrm{H_{3}OF}\), \(\mathrm{H_{4}}\), \(\mathrm{H_{4}O_{2}}\), \(\mathrm{H_{5}NO}\), \(\mathrm{HF}\), \(\mathrm{HNO}\), \(\mathrm{HNO_{2}}\), \(\mathrm{Li_{2}O}\), \(\mathrm{Li_{2}O_{2}}\), \(\mathrm{LiBH_{4}}\), \(\mathrm{LiF}\), \(\mathrm{LiNH_{2}}\), \(\mathrm{LiNH_{2}O}\), \(\mathrm{LiOF}\), \(\mathrm{LiOH}\), \(\mathrm{N_{2}}\), \(\mathrm{N_{2}H_{2}}\), \(\mathrm{N_{2}H_{4}}\), \(\mathrm{N_{2}O}\), \(\mathrm{NH_{2}F}\), \(\mathrm{NH_{3}}\), \(\mathrm{NH_{3}O}\), \(\mathrm{NO}\), \(\mathrm{O_{3}}\)~\cite{lac_curriculum,cccbdb,ptcomp} \\
\addlinespace[4pt]
Third-period-containing elements & 22 & 10,000 & 12,499 & \(\mathrm{Al}\), \(\mathrm{CS_{2}}\), \(\mathrm{Cl_{2}}\), \(\mathrm{H_{2}S}\), \(\mathrm{HCl}\), \(\mathrm{LiCl}\), \(\mathrm{Mg}\), \(\mathrm{MgO}\), \(\mathrm{NaCl}\), \(\mathrm{NaF}\), \(\mathrm{NaH}\), \(\mathrm{P}\), \(\mathrm{PF_{3}}\), \(\mathrm{S}\), \(\mathrm{SO_{2}}\), \(\mathrm{SO_{3}}\), \(\mathrm{Si}\), \(\mathrm{SiCl_{4}}\), \(\mathrm{SiF_{4}}\), \(\mathrm{SiH_{4}}\), \(\mathrm{SiO}\), \(\mathrm{SiS}\)~\cite{cccbdb,ptcomp,agar1962,edwards1967} \\
\addlinespace[4pt]
Fourth-period-containing elements & 50 & 10,000 & 12,501 & \(\mathrm{AsF_{3}}\), \(\mathrm{AsH_{3}}\), \(\mathrm{Br}\), \(\mathrm{Ca}\), \(\mathrm{CaO}\), \(\mathrm{Co}\), \(\mathrm{CoCl}\), \(\mathrm{CoH}\), \(\mathrm{CoO}\), \(\mathrm{CoS}\), \(\mathrm{CrF}\), \(\mathrm{CrO}\), \(\mathrm{CuCl}\), \(\mathrm{CuO}\), \(\mathrm{CuS}\), \(\mathrm{FeCl}\), \(\mathrm{FeH}\), \(\mathrm{FeS}\), \(\mathrm{Ga}\), \(\mathrm{Ge}\), \(\mathrm{HBr}\), \(\mathrm{K}\), \(\mathrm{KBr}\), \(\mathrm{KCl}\), \(\mathrm{KF}\), \(\mathrm{KH}\), \(\mathrm{LiBr}\), \(\mathrm{Mn}\), \(\mathrm{MnO}\), \(\mathrm{MnS}\), \(\mathrm{Ni}\), \(\mathrm{NiCl}\), \(\mathrm{NiF}\), \(\mathrm{NiO}\), \(\mathrm{ScH}\), \(\mathrm{ScO}\), \(\mathrm{ScS}\), \(\mathrm{Se}\), \(\mathrm{TiCl}\), \(\mathrm{TiH}\), \(\mathrm{TiN}\), \(\mathrm{TiS}\), \(\mathrm{V}\), \(\mathrm{VCl}\), \(\mathrm{VN}\), \(\mathrm{VO}\), \(\mathrm{Zn}\), \(\mathrm{ZnCl}\), \(\mathrm{ZnH}\), \(\mathrm{ZnS}\)~\cite{cccbdb,aoto2017,ptcomp} \\
\addlinespace[4pt]
\midrule
\textbf{Total} & \textbf{137} & \textbf{25,000} & \textbf{30,270} & \\
\bottomrule
\end{tabularx}

\smallskip
\begin{minipage}{\linewidth}
\footnotesize
Source configurations include published LAC scans, reference molecular structures, and isolated atoms. Additional spin sectors share nuclear coordinates. Per-system descriptions and geometry sources are provided in \hyperref[app:appendix-3]{Appendix 4}.
\end{minipage}
\end{table}

Pretraining used 1 molecule per NVIDIA RTX PRO 6000 Blackwell graphics processing unit (GPU). 32 of the same GPUs were used with 256 walkers per
molecule for a total of 8,192 walkers. Adaptation uses 1,024 walkers on one GPU for its target system. Table~\ref{tab:model-comparison} compares the architecture and protocol with SAAM and Orbformer.

The pretrained model was initialized randomly and without HF fitting. Warm adaptations initialized neural parameters from the specified pretrained checkpoint, with fresh optimizer and walker states. Random controls used freshly initialized neural parameters. HF controls underwent 2,000 orbital-fitting updates before VMC, using aug-cc-pVDZ for all-electron targets and ccECP-cc-pVTZ for ccECP targets.

Electron configurations were sampled using random-walk Metropolis.
Random-walk Metropolis uses displacement proposals with an acceptance-rejection step~\cite{metropolis1953,hastings1970}. 

All energy-comparing runs use 16 determinant spatial components and a linear learning-rate warmup of
\(10^{-7}\to10^{-3}\) over 500 updates. After the 500-update warmup, the learning rate follows an
inverse-time scale of 10,000 updates. We use KFAC damping of \(10^{-3}\), KFAC norm constraint of \(10^{-3}\),
curvature moving-average decay of 0.95, median-centered local-energy clipping of \(5\times\) mean absolute
deviation from median for optimization, 12 point ccECP angular quadrature, and 32-bit floating-point (FP32) parameters and local energies.

\subsection{Independent energy evaluation}

For the transfer-energy comparisons in Figures~\ref{fig:observable-accuracy}
and~\ref{fig:component-transfer}, all learned parameters are held fixed after 20,000 adaptation updates. Each adaptation seed is evaluated using two separately seeded persistent Metropolis chains initialized from saved walkers. Each uses 256 walkers, 1,200 burn-in sweeps, and 512 measurement records separated by 60 sweeps. Reported energies combine the FP32 mean with a paired FP64 correction evaluated on identical configurations every sixteenth record (\hyperref[app:appendix-2]{Appendix 2}). 

All sampled local energies are retained. Time-blocking accounts for serially correlated measurements~\cite{flyvbjerg1989}. For each checkpoint, we use the largest uncertainty estimate from the time-blocking and between-chain variation, adding the uncertainty in the FP64-FP32 correction. Component uncertainties are propagated to energy differences. These estimates exclude optimization and Hamiltonian approximation errors.

\subsection{Observables and reference conventions}

We evaluate bond dissociation energies, spin-state gaps, ionization energies, and reaction energy from their component
energies. Electronic dissociation energies, \(D_e\), are calculated as the sum of the separated-atom
energies minus the molecular energy. Spin-state gaps are the energy of the first state named minus the second;
for O\textsubscript{2} and Fe\textsuperscript{+}, these are singlet minus triplet and quartet minus
sextet, respectively. Ionization energies are calculated as cation minus neutral-atom energies.

Atomic spectroscopic reference data are available from the NIST Atomic Spectra Database~\cite{ref13}
and the NIST Handbook~\cite{nist_handbook}. The tabulated Fe\textsuperscript{+}, Cr, C, N, and O spin-gap
references are reproduced, to the shown precision, by differences of term centroids
\(\bar E=\sum_J(2J+1)E_J/\sum_J(2J+1)\) using the Handbook levels.
The CuH dissociation reference is the conversion of
\(D_e=22\,630\pm250\,\mathrm{cm}^{-1}\) reported by Mohammadian and Shayesteh~\cite{ref10}.
Electronic well depths \(D_e\), zero-point dissociation energies \(D_0\), and finite-temperature
bond dissociation enthalpies must be distinguished when using experimental thermochemistry
and spectroscopic constants~\cite{ref15,ref16,ref17}.

\section{Acknowledgements}

We thank Hanggai Nuomin for reviews of earlier drafts. We thank Adam Foster for advice on training the Orbformer architecture and suggestions on the training curriculum and model evaluations.

\section{Funding}

We acknowledge funding from Primateria.

\bibliography{references}

\clearpage
\section{Appendix 1: Forward Laplacian}
\label{app:appendix-1}

Forward Laplacian calculates the derivatives of a neural network's intermediate features
\(\Delta\mathbf h^l\) and output \(y=\log|\Psi_\theta|\) with respect to the electron positions \(x\)
avoiding constructing the full Hessian. Instead, \(\Delta y\) is computed directly using the first
derivatives and the intermediate Laplacian term~\cite{ref9}.
\[
\begin{bmatrix}\mathbf x\\\nabla\mathbf x\\\Delta\mathbf x\end{bmatrix}
\to\cdots\to
\begin{bmatrix}\mathbf h^l\\\nabla\mathbf h^l\\\Delta\mathbf h^l\end{bmatrix}
\to\cdots\to
\begin{bmatrix}y\\\nabla y\\\Delta y\end{bmatrix}
\]
The table compares loop-based and forward-Laplacian evaluation times.

\begingroup
\small\setlength{\tabcolsep}{5pt}\renewcommand{\arraystretch}{1.3}
\noindent\begin{tabularx}{\linewidth}{@{}lYYY@{}}
\toprule
\textbf{System} &  \textbf{Loop time (s)} & \textbf{Forward time (s)} & \textbf{Time reduction}\\
\midrule
CuH & 3.707 & 2.76 & 25.55\%\\
ScCl & 4.898 & 3.995 & 18.44\%\\
\bottomrule
\end{tabularx}
\endgroup

We compare loop-based and forward-Laplacian local-energy evaluation on identical fixed electron
configurations using FP64 arithmetic. Timings correspond to a 32-configuration batch, rather than a
complete optimization update.

\section{Appendix 2: FP64 Correction}
\label{app:appendix-2}

We evaluate frozen checkpoints primarily in FP32 to reduce computational cost and optimize calculations for NVIDIA RTX PRO 6000 Blackwells. Every 16th measurement record is also evaluated in FP64 using the same electron configurations and unchanged model parameters. The mean
difference between these paired evaluations corrects the FP32 energy:
\[
E_{\mathrm{reported}}
=
\overline{E}_{\mathrm{FP32}}
+
\overline{
E_{\mathrm{FP64}}-E_{\mathrm{FP32}}
}_{\mathrm{paired}}.
\]

Statistical uncertainties account for correlations between successive measurements and differences between the two sampling streams. The uncertainty in the precision correction is added to the FP32 uncertainty.

\begin{landscape}
\section{Appendix 3: Finetuning vs Cold Training Results}

\begingroup
\fontsize{10}{12}\selectfont
\setlength{\tabcolsep}{3pt}\renewcommand{\arraystretch}{1.15}
\setlength{\LTleft}{0pt}\setlength{\LTright}{0pt plus 1fil}
\begin{longtable}{@{}L{5cm}C{0.8cm}C{3.25cm}C{3.25cm}C{3.25cm}C{3.25cm}C{\dimexpr\linewidth-18.8cm-12\tabcolsep\relax}@{}}
\caption{Frozen component energies and derived observables after 20,000 accepted VMC adaptation updates for cold starts and the 100k pretrained checkpoint.}\label{tab:transfer-results}\\
\toprule
 & & \multicolumn{2}{c}{\textbf{Cold start (two-seed means)}} & \multicolumn{2}{c}{\textbf{Pretrained (100k)}} & \\
\cmidrule(lr){3-4}\cmidrule(lr){5-6}
\makecell[tl]{\textbf{Observable /}\\\textbf{component}} & \textbf{Unit} & \makecell[t]{\textbf{HF-}\\\textbf{initialized}} & \makecell[t]{\textbf{Random-}\\\textbf{initialized}} & \textbf{Seed 42} & \textbf{Seed 43} & \textbf{Reference}\\
\midrule\endfirsthead
\toprule
 & & \multicolumn{2}{c}{\textbf{Cold start (two-seed means)}} & \multicolumn{2}{c}{\textbf{Pretrained (100k)}} & \\
\cmidrule(lr){3-4}\cmidrule(lr){5-6}
\makecell[tl]{\textbf{Observable /}\\\textbf{component}} & \textbf{Unit} & \makecell[t]{\textbf{HF-}\\\textbf{initialized}} & \makecell[t]{\textbf{Random-}\\\textbf{initialized}} & \textbf{Seed 42} & \textbf{Seed 43} & \textbf{Reference}\\
\midrule\endhead
\midrule\multicolumn{7}{r}{\textit{Continued on next page}}\\\endfoot
\bottomrule\endlastfoot

\rowcolor{gray!13}O\textsubscript{2} singlet-triplet gap & mHa & \valerr{-45.47}{62.75}\textsuperscript{$\dagger$} & \valerr{48.37}{4.04} & \valerr{51.17}{4.59} & \valerr{47.46}{6.63} & \(\approx36.08\)~\cite{oxygen_nist} \\* \addlinespace[2pt]
O\textsubscript{2} singlet & Ha & \valerr{-150.266430}{0.003778} & \valerr{-150.242842}{0.002812} & \valerr{-150.250238}{0.002155} & \valerr{-150.246275}{0.005929} & \\* \addlinespace[2pt]
O\textsubscript{2} triplet & Ha & \valerr{-150.220958}{0.066531}\textsuperscript{$\dagger$} & \valerr{-150.291213}{0.004973} & \valerr{-150.301408}{0.004057} & \valerr{-150.293736}{0.002963} & \\ \addlinespace[2pt]

\rowcolor{gray!13}Fe\textsuperscript{+} quartet-sextet gap & mHa & \valerr{11.11}{11.51} & \valerr{108.08}{4.05} & \valerr{6.86}{4.47} & \valerr{17.19}{5.11} & \(9.11\)~\cite{nist_handbook} \\* \addlinespace[2pt]
Fe\textsuperscript{+} quartet & Ha & \valerr{-123.051085}{0.004167} & \valerr{-122.945875}{0.003104} & \valerr{-123.054808}{0.002500} & \valerr{-123.049157}{0.004479} & \\* \addlinespace[2pt]
Fe\textsuperscript{+} sextet & Ha & \valerr{-123.062200}{0.010726} & \valerr{-123.053951}{0.002607} & \valerr{-123.061668}{0.003703} & \valerr{-123.066343}{0.002455} & \\ \addlinespace[2pt]

\rowcolor{gray!13}Cr quintet-septet gap & mHa & \valerr{41.63}{4.81} & \valerr{-89.98}{6.57} & \valerr{40.49}{3.13} & \valerr{43.36}{4.25} & \(34.60\)~\cite{nist_handbook} \\* \addlinespace[2pt]
Cr quintet & Ha & \valerr{-86.573407}{0.003940} & \valerr{-86.559450}{0.003283} & \valerr{-86.571034}{0.001886} & \valerr{-86.567494}{0.001830} & \\* \addlinespace[2pt]
Cr septet & Ha & \valerr{-86.615034}{0.002755} & \valerr{-86.469468}{0.004171} & \valerr{-86.611526}{0.002503} & \valerr{-86.610851}{0.003841} & \\ \addlinespace[2pt]

\rowcolor{gray!13}C singlet-triplet gap & mHa & \valerr{66.34}{1.47} & \valerr{50.08}{1.64} & \valerr{38.56}{5.19} & \valerr{29.68}{12.19} & \(46.31\)~\cite{nist_handbook} \\* \addlinespace[2pt]
C singlet & Ha & \valerr{-37.771769}{0.001126} & \valerr{-37.788214}{0.001247} & \valerr{-37.795344}{0.001003} & \valerr{-37.796961}{0.001133} & \\* \addlinespace[2pt]
C triplet & Ha & \valerr{-37.838108}{0.000949} & \valerr{-37.838289}{0.001067} & \valerr{-37.833908}{0.005089} & \valerr{-37.826644}{0.012133} & \\ \addlinespace[2pt]

\rowcolor{gray!13}N doublet-quartet gap & mHa & \valerr{112.35}{1.81} & \valerr{95.53}{6.16} & \valerr{88.23}{2.04} & \valerr{88.12}{2.43} & \(87.61\)~\cite{nist_handbook} \\* \addlinespace[2pt]
N doublet & Ha & \valerr{-54.472810}{0.001242} & \valerr{-54.485102}{0.008520} & \valerr{-54.496087}{0.001438} & \valerr{-54.496491}{0.001312} & \\* \addlinespace[2pt]
N quartet & Ha & \valerr{-54.585155}{0.001642} & \valerr{-54.580635}{0.002358} & \valerr{-54.584313}{0.001452} & \valerr{-54.584614}{0.002048} & \\ \addlinespace[2pt]

\rowcolor{gray!13}O singlet-triplet gap & mHa & \valerr{88.17}{13.61} & \valerr{-168.50}{239.32} & \valerr{75.08}{2.34} & \valerr{75.01}{2.84} & \(71.94\)~\cite{nist_handbook} \\* \addlinespace[2pt]
O singlet & Ha & \valerr{-74.973164}{0.015153} & \valerr{-74.984847}{0.001592} & \valerr{-74.986534}{0.001525} & \valerr{-74.986999}{0.001673} & \\* \addlinespace[2pt]
O triplet & Ha & \valerr{-75.061337}{0.001543} & \valerr{-74.816352}{0.239148} & \valerr{-75.061609}{0.001773} & \valerr{-75.062013}{0.002291} & \\ \addlinespace[2pt]

\rowcolor{gray!13}CuH binding, $D_e$ & mHa & \valerr{76.62}{6.99} & \valerr{68.68}{83.65} & \valerr{112.15}{5.92} & \valerr{105.25}{4.93} & \(103.11\pm1.14\)~\cite{ref10} \\* \addlinespace[2pt]
CuH singlet & Ha & \valerr{-196.940475}{0.004709} & \valerr{-196.891077}{0.062137} & \valerr{-196.960978}{0.003782} & \valerr{-196.964404}{0.003892} & \\* \addlinespace[2pt]
Cu doublet & Ha & \valerr{-196.363854}{0.004701} & \valerr{-196.322394}{0.021516} & \valerr{-196.348825}{0.004561} & \valerr{-196.359153}{0.003023} & \\ \addlinespace[2pt]

\rowcolor{gray!13}CrH binding, $D_e$ & mHa & \valerr{69.08}{4.63} & \valerr{162.06}{10.40} & \valerr{80.34}{6.87} & \valerr{86.83}{5.29} & \(81.51\pm2.67\)~\cite{ref11} \\* \addlinespace[2pt]
CrH sextet & Ha & \valerr{-87.184114}{0.003724} & \valerr{-87.131533}{0.014574} & \valerr{-87.191863}{0.006392} & \valerr{-87.197682}{0.003639} & \\* \addlinespace[2pt]
Cr septet & Ha & \valerr{-86.615034}{0.002755} & \valerr{-86.469468}{0.004171} & \valerr{-86.611526}{0.002503} & \valerr{-86.610851}{0.003841} & \\ \addlinespace[2pt]

\rowcolor{gray!13}Sc first ionization & mHa & \valerr{239.52}{2.18} & \valerr{179.15}{55.40} & \valerr{233.06}{2.25} & \valerr{232.67}{1.97} & \(241.15\)~\cite{nist_handbook} \\* \addlinespace[2pt]
Sc\textsuperscript{+} triplet & Ha & \valerr{-46.303001}{0.001236} & \valerr{-46.295997}{0.002268} & \valerr{-46.304802}{0.001547} & \valerr{-46.305821}{0.001417} & \\* \addlinespace[2pt]
Sc doublet & Ha & \valerr{-46.542518}{0.001427} & \valerr{-46.475146}{0.056545} & \valerr{-46.537859}{0.001639} & \valerr{-46.538490}{0.001363} & \\ \addlinespace[2pt]

\rowcolor{gray!13}Ti first ionization & mHa & \valerr{242.58}{3.31} & \valerr{236.49}{3.19} & \valerr{241.24}{2.44} & \valerr{243.32}{2.40} & \(250.94\)~\cite{nist_handbook} \\* \addlinespace[2pt]
Ti\textsuperscript{+} quartet & Ha & \valerr{-57.825617}{0.002190} & \valerr{-57.818322}{0.001831} & \valerr{-57.828593}{0.001924} & \valerr{-57.827985}{0.001784} & \\* \addlinespace[2pt]
Ti triplet & Ha & \valerr{-58.068199}{0.001373} & \valerr{-58.054815}{0.003975} & \valerr{-58.069831}{0.001506} & \valerr{-58.071309}{0.001611} & \\ \addlinespace[2pt]

\rowcolor{gray!13}Cr first ionization & mHa & \valerr{275.20}{28.07} & \valerr{168.15}{6.76} & \valerr{243.35}{3.59} & \valerr{245.12}{4.56} & \(248.66\)~\cite{nist_handbook} \\* \addlinespace[2pt]
Cr\textsuperscript{+} sextet & Ha & \valerr{-86.339835}{0.027294} & \valerr{-86.301322}{0.003282} & \valerr{-86.368174}{0.002567} & \valerr{-86.365729}{0.002465} & \\* \addlinespace[2pt]
Cr septet & Ha & \valerr{-86.615034}{0.002755} & \valerr{-86.469468}{0.004171} & \valerr{-86.611526}{0.002503} & \valerr{-86.610851}{0.003841} & \\ \addlinespace[2pt]

\rowcolor{gray!13}Cu + H\textsubscript{2} $\to$ CuH + H & mHa & \valerr{95.48}{7.08} & \valerr{103.26}{83.60} & \valerr{61.82}{5.93} & \valerr{68.31}{4.94} & \(71.39\pm1.14\)~\cite{ref10,ref15,irikura2007} \\* \addlinespace[2pt]
CuH singlet & Ha & \valerr{-196.940475}{0.004709} & \valerr{-196.891077}{0.062137} & \valerr{-196.960978}{0.003782} & \valerr{-196.964404}{0.003892} & \\* \addlinespace[2pt]
Cu doublet & Ha & \valerr{-196.363854}{0.004701} & \valerr{-196.322394}{0.021516} & \valerr{-196.348825}{0.004561} & \valerr{-196.359153}{0.003023} & \\* \addlinespace[2pt]
H\textsubscript{2} singlet & Ha & \valerr{-1.172098}{0.000381} & \valerr{-1.171943}{0.000390} & \valerr{-1.173972}{0.000243} & \valerr{-1.173561}{0.000336} & \\

\end{longtable}

\par\smallskip\footnotesize
HF starts include 2,000 orbital-fitting updates before VMC. All columns use the same architecture.
Cold columns average adaptation seeds 42 and 43; uncertainties use the larger of the propagated Monte Carlo and between-seed standard errors. Observables are calculated within each seed before averaging. Pretrained-column uncertainties are Monte Carlo standard errors for individual runs. Independent component uncertainties are propagated in quadrature; $E(\mathrm H)=-0.5$ Ha. All frozen measurements use persistent Metropolis chains. $\dagger$ The HF O$_2$ triplet includes a read with substantial energy drift during frozen measurement.
\par\smallskip
\endgroup\end{landscape}

\section{Appendix 4: Curriculum Provenance}
\label{app:appendix-3}

\begingroup
\scriptsize
\setlength{\tabcolsep}{3.5pt}
\renewcommand{\arraystretch}{1.13}
\setlength{\LTleft}{0pt}\setlength{\LTright}{0pt}
\begin{longtable}{@{}>{\raggedright\arraybackslash}p{0.20\linewidth}
>{\centering\arraybackslash}p{0.055\linewidth}
>{\centering\arraybackslash}p{0.06\linewidth}
>{\raggedright\arraybackslash}p{0.24\linewidth}
>{\centering\arraybackslash}p{0.065\linewidth}
>{\centering\arraybackslash}p{0.07\linewidth}
>{\raggedright\arraybackslash}p{0.21\linewidth}@{}}
\caption{Geometry sources and prescribed training spins for all 137 compositions
in the pretraining curriculum. Ng counts nuclear geometries, including generated
distortions; NgS counts geometry-spin configurations. Distances are in \AA.
}

\label{tab:curriculum-v5-provenance}\\
\toprule
Composition(s) & $N_g$ & $N_{gS}$ & Parent geometry / configurations & Training $S$ & Reference & Notes \\
\midrule\endfirsthead
\multicolumn{7}{l}{\tablename\ \thetable{} --- continued}\\
\toprule
Composition(s) & $N_g$ & $N_{gS}$ & Parent geometry / configurations & Training $S$ & Reference & Notes \\
\midrule\endhead
\midrule\multicolumn{7}{r}{Continued on next page}\\\endfoot
\bottomrule\endlastfoot
\(\mathrm{Al}\) & 1 & 1 & Isolated atom at origin & \(1/2\) & \cite{ptcomp} &  \\
\(\mathrm{AsF_{3}}\) & 257 & 322 & \(r_g(\mathrm{AsF})=1.706\); \(\angle\mathrm{FAsF}=96.2^\circ\) & \(0, 1\) & \cite{cccbdb} & Electron-diffraction \(r_g\)\\
\(\mathrm{AsH_{3}}\) & 257 & 322 & \(r(\mathrm{AsH})=1.51084\); \(\angle\mathrm{HAsH}=92.08179^\circ\) & \(0, 1\) & \cite{cccbdb} &  \\
\(\mathrm{B}\) & 1 & 1 & Isolated atom at origin & \(1/2\) & \cite{ptcomp} &  \\
\(\mathrm{B_{2}CH_{2}}\), \(\mathrm{BCH_{2}F}\), \(\mathrm{BCH_{3}}\), \(\mathrm{BCH_{3}O}\), \(\mathrm{BCH_{4}N}\), \(\mathrm{BH_{2}F}\), \(\mathrm{BH_{3}}\), \(\mathrm{BH_{3}O_{2}}\), \(\mathrm{C_{2}H_{2}}\), \(\mathrm{C_{2}H_{3}Li}\), \(\mathrm{C_{2}H_{4}}\), \(\mathrm{C_{2}H_{4}O}\), \(\mathrm{C_{2}H_{5}Li}\), \(\mathrm{C_{2}H_{6}}\), \(\mathrm{C_{2}Li_{2}}\), \(\mathrm{CH_{2}FLi}\), \(\mathrm{CH_{3}F}\), \(\mathrm{CHF}\), \(\mathrm{CHFO}\), \(\mathrm{LiBH_{4}}\) & 85 & 85 & LAC & \(0\) & \cite{lac_curriculum} &  \\
\(\mathrm{BF_{3}}\) & 1 & 1 & \(r(\mathrm{BF})=1.307\); \(\angle\mathrm{FBF}=119.99989^\circ\) & \(0\) & \cite{cccbdb} &  \\
\(\mathrm{Be}\) & 1 & 1 & Isolated atom at origin & \(0\) & \cite{ptcomp} &  \\
\(\mathrm{Br}\) & 1 & 1 & Isolated atom at origin & \(1/2\) & \cite{ptcomp} &  \\
\(\mathrm{C_{2}}\) & 85 & 255 & LAC; \(r=0.72\text{--}21.33935\) & \(0, 1, 2\) & \cite{lac_curriculum} &  \\
\(\mathrm{C_{2}H_{3}F}\) & 46 & 46 & LAC; corrected vinyl-fluoride identity & \(0\) & \cite{lac_curriculum} \\
\(\mathrm{C_{2}H_{3}N}\) & 39 & 39 & LAC & \(0\) & \cite{lac_curriculum} &  \\
\(\mathrm{CFN}\), \(\mathrm{CH_{2}O}\), \(\mathrm{CH_{2}O_{2}}\), \(\mathrm{CH_{3}Li}\), \(\mathrm{CH_{3}N}\), \(\mathrm{CH_{3}NO}\), \(\mathrm{CH_{4}LiN}\), \(\mathrm{CH_{4}O}\), \(\mathrm{CH_{5}N}\), \(\mathrm{CHLiO}\), \(\mathrm{CHN}\), \(\mathrm{CLiN}\), \(\mathrm{H_{3}OF}\), \(\mathrm{H_{4}O_{2}}\), \(\mathrm{H_{5}NO}\), \(\mathrm{HNO}\), \(\mathrm{HNO_{2}}\), \(\mathrm{Li_{2}O}\), \(\mathrm{Li_{2}O_{2}}\), \(\mathrm{LiNH_{2}}\), \(\mathrm{LiNH_{2}O}\), \(\mathrm{LiOF}\), \(\mathrm{LiOH}\), \(\mathrm{N_{2}H_{2}}\), \(\mathrm{N_{2}H_{4}}\), \(\mathrm{N_{2}O}\), \(\mathrm{NH_{2}F}\), \(\mathrm{NH_{3}O}\), \(\mathrm{O_{3}}\) & 84 & 84 & LAC & \(0\) & \cite{lac_curriculum} &  \\
\(\mathrm{CO}\) & 84 & 84 & LAC; \(r=0.6768\text{--}20.05899\) & \(0\) & \cite{lac_curriculum} &  \\
\(\mathrm{CS_{2}}\) & 588 & 735 & Linear; \(r_e(\mathrm{CS})=1.5529\) & \(0, 1\) & \cite{agar1962}\\
\(\mathrm{Ca}\) & 1 & 1 & Isolated atom at origin & \(0\) & \cite{ptcomp} &  \\
\(\mathrm{CaO}\) & 256 & 320 & \(r_e=1.8221\) & \(0, 1\) & \cite{cccbdb} &  \\
\(\mathrm{Cl_{2}}\) & 588 & 735 & \(r_e=1.9879\) & \(0, 1\) & \cite{cccbdb} &  \\
\(\mathrm{Co}\) & 1 & 1 & Isolated atom at origin & \(3/2\) & \cite{ptcomp} &  \\
\(\mathrm{CoCl}\) & 256 & 320 & \(r_e=2.0656\) & \(1, 2\) & \cite{aoto2017} &  \\
\(\mathrm{CoH}\) & 256 & 320 & \(r_e=1.5327\) & \(1, 2\) & \cite{aoto2017} &  \\
\(\mathrm{CoO}\) & 256 & 320 & \(r_e=1.6286\) & \(3/2, 5/2\) & \cite{aoto2017} &  \\
\(\mathrm{CoS}\) & 256 & 320 & \(r_e=1.9786\) & \(3/2, 5/2\) & \cite{aoto2017} &  \\
\(\mathrm{CrF}\) & 256 & 320 & \(r_e=1.7839\) & \(5/2, 7/2\) & \cite{aoto2017} &  \\
\(\mathrm{CrO}\) & 256 & 320 & \(r_e=1.615\) & \(2, 3\) & \cite{aoto2017} &  \\
\(\mathrm{CuCl}\) & 256 & 320 & \(r_e=2.0512\) & \(0, 1\) & \cite{aoto2017} &  \\
\(\mathrm{CuO}\) & 256 & 320 & \(r_e=1.7246\); \newline \(r_e=1.7243\) & \(1/2, 3/2\) & \cite{aoto2017,cccbdb} &  \\
\(\mathrm{CuS}\) & 256 & 320 & \(r_e=2.0499\) & \(1/2, 3/2\) & \cite{aoto2017} &  \\
\(\mathrm{F}\) & 1 & 1 & Isolated atom at origin & \(1/2\) & \cite{ptcomp} &  \\
\(\mathrm{F_{2}}\) & 85 & 85 & 84 LAC; \(r=0.88356\text{--}25.25157\); \newline \(r_e=1.4119\) & \(0\) & \cite{lac_curriculum,cccbdb} &  \\
\(\mathrm{FeCl}\) & 256 & 320 & \(r_e=2.1742\) & \(5/2, 7/2\) & \cite{aoto2017} &  \\
\(\mathrm{FeH}\) & 256 & 320 & \(r_e=1.606\) & \(3/2, 5/2\) & \cite{aoto2017} &  \\
\(\mathrm{FeS}\) & 256 & 320 & \(r_e=2.0140\), inferred using \(B_e=(3B_0-B_1)/2\) & \(2, 3\) & \cite{aoto2017, takano2004} \\
\(\mathrm{Ga}\) & 1 & 1 & Isolated atom at origin & \(1/2\) & \cite{ptcomp} &  \\
\(\mathrm{Ge}\) & 1 & 1 & Isolated atom at origin & \(1\) & \cite{ptcomp} &  \\
\(\mathrm{H_{2}F_{2}}\) & 49 & 49 & LAC & \(0\) & \cite{lac_curriculum} &  \\
\(\mathrm{H_{2}S}\) & 588 & 735 & \(r_e(\mathrm{SH})=1.3356\); \(\angle\mathrm{HSH}=92^\circ07^\prime\) & \(0, 1\) & \cite{edwards1967} & \\
\(\mathrm{H_{4}}\) & 50 & 150 & LAC & \(0, 1, 2\) & \cite{lac_curriculum} &  \\
\(\mathrm{HBr}\) & 257 & 322 & \(r_e=1.4144\) & \(0, 1\) & \cite{cccbdb} &  \\
\(\mathrm{HCl}\) & 588 & 735 & \(r_e=1.2746\) & \(0, 1\) & \cite{cccbdb} &  \\
\(\mathrm{HF}\) & 84 & 84 & LAC; \(r=0.5502\text{--}16.30682\) & \(0\) & \cite{lac_curriculum} &  \\
\(\mathrm{K}\) & 1 & 1 & Isolated atom at origin & \(1/2\) & \cite{ptcomp} &  \\
\(\mathrm{KBr}\) & 257 & 322 & \(r_e=2.8208\) & \(0, 1\) & \cite{cccbdb} &  \\
\(\mathrm{KCl}\) & 256 & 320 & \(r_e=2.6667\) & \(0, 1\) & \cite{cccbdb} &  \\
\(\mathrm{KF}\) & 256 & 320 & \(r_e=2.1715\) & \(0, 1\) & \cite{cccbdb} &  \\
\(\mathrm{KH}\) & 256 & 320 & \(r_e=2.2425\) & \(0, 1\) & \cite{cccbdb} &  \\
\(\mathrm{LiBr}\) & 257 & 322 & \(r_e=2.1704\) & \(0, 1\) & \cite{cccbdb} \\
\(\mathrm{LiCl}\) & 588 & 735 & \(r_e=2.0207\) & \(0, 1\) & \cite{cccbdb} &  \\
\(\mathrm{LiF}\) & 85 & 85 & 84 LAC; \(r=0.94439\text{--}27.98976\); \newline \(r_e=1.5639\) & \(0\) & \cite{lac_curriculum,cccbdb} &  \\
\(\mathrm{Mg}\) & 1 & 1 & Isolated atom at origin & \(0\) & \cite{ptcomp} &  \\
\(\mathrm{MgO}\) & 588 & 735 & \(r_e=1.749\) & \(0, 1\) & \cite{cccbdb} &  \\
\(\mathrm{Mn}\) & 1 & 1 & Isolated atom at origin & \(5/2\) & \cite{ptcomp} &  \\
\(\mathrm{MnO}\) & 256 & 320 & \(r_e=1.6446\); Pekeris/Morse inference & \(5/2, 7/2\) & \cite{aoto2017}\\
\(\mathrm{MnS}\) & 256 & 320 & \(r_e=2.0663\) & \(5/2, 7/2\) & \cite{aoto2017} &  \\
\(\mathrm{N_{2}}\) & 84 & 84 & LAC; \(r=0.65856\text{--}19.51839\) & \(0\) & \cite{lac_curriculum} &  \\
\(\mathrm{NH_{3}}\) & 85 & 85 & 84 LAC; \newline Historical \(r_e(\mathrm{NH})=1.0124\); \(\angle\mathrm{HNH}=106.67^\circ\); rounded XYZ retained & \(0\) & \cite{lac_curriculum,cccbdb} \\
\(\mathrm{NO}\) & 84 & 84 & LAC; \(r=0.69\text{--}20.45021\) & \(1/2\) & \cite{lac_curriculum} &  \\
\(\mathrm{NaCl}\) & 588 & 735 & \(r_e=2.3608\) & \(0, 1\) & \cite{cccbdb} &  \\
\(\mathrm{NaF}\) & 588 & 735 & \(r_e=1.9259\) & \(0, 1\) & \cite{cccbdb} &  \\
\(\mathrm{NaH}\) & 588 & 735 & \(r_e=1.8874\) & \(0, 1\) & \cite{cccbdb} &  \\
\(\mathrm{Ni}\) & 1 & 1 & Isolated atom at origin & \(1\) & \cite{ptcomp} &  \\
\(\mathrm{NiCl}\) & 256 & 320 & \(r_e=2.0615\); Pekeris/Morse inference & \(1/2, 3/2\) & \cite{aoto2017, hirao2000} & Reconstructed from original constants; printed source distance inconsistent \\
\(\mathrm{NiF}\) & 256 & 320 & \(r_e=1.7387\) & \(1/2, 3/2\) & \cite{aoto2017} &  \\
\(\mathrm{NiO}\) & 256 & 320 & \(r_e=1.6271\) & \(1, 2\) & \cite{aoto2017} &  \\
\(\mathrm{P}\) & 1 & 1 & Isolated atom at origin & \(3/2\) & \cite{ptcomp} &  \\
\(\mathrm{PF_{3}}\) & 588 & 735 & \(r_e(\mathrm{PF})=1.561\); \(\angle\mathrm{FPF}=97.7^\circ\); reconstructed from internals & \(0, 1\) & \cite{cccbdb} \\
\(\mathrm{S}\) & 1 & 1 & Isolated atom at origin & \(1\) & \cite{ptcomp} &  \\
\(\mathrm{SO_{2}}\) & 588 & 735 &
Effective \(r_0(\mathrm{SO})\simeq1.4321\);
\(\angle\mathrm{OSO}\simeq119.5^\circ\) &
\(0, 1\) & \cite{cccbdb,herzberg1966} & \\
\(\mathrm{SO_{3}}\) & 588 & 735 & \(r_e(\mathrm{SO})=1.4175\); planar \(D_{3h}\), \(120^\circ\) & \(0, 1\) & \cite{cccbdb} & Inferred equilibrium structure \\
\(\mathrm{ScH}\) & 256 & 320 & \(r_e=1.7754\) & \(0, 1\) & \cite{aoto2017} &  \\
\(\mathrm{ScO}\) & 256 & 320 & \(r_e=1.6661\) & \(1/2, 3/2\) & \cite{aoto2017} &  \\
\(\mathrm{ScS}\) & 256 & 320 & \(r_e=2.1353\) & \(1/2, 3/2\) & \cite{aoto2017} &  \\
\(\mathrm{Se}\) & 1 & 1 & Isolated atom at origin & \(1\) & \cite{ptcomp} &  \\
\(\mathrm{Si}\) & 1 & 1 & Isolated atom at origin & \(1\) & \cite{ptcomp} &  \\
\(\mathrm{SiCl_{4}}\) & 588 & 735 & \(r_a(\mathrm{SiCl})=2.019\) at 251 K; tetrahedral & \(0, 1\) & \cite{cccbdb} \\
\(\mathrm{SiF_{4}}\) & 588 & 735 &
Effective \(r_0(\mathrm{SiF})\simeq1.554\); tetrahedral &
\(0, 1\) & \cite{cccbdb,wang2000} & \\
\(\mathrm{SiH_{4}}\) & 588 & 735 & Effective ground-state SiH distance \(\simeq1.47986\) in retained XYZ; source \(1.4798\pm0.0004\) & \(0, 1\) & \cite{cccbdb} & Assumes unchanged structure across isotopes \\
\(\mathrm{SiO}\) & 588 & 735 & \(r_e=1.5097\) & \(0, 1\) & \cite{cccbdb} &  \\
\(\mathrm{SiS}\) & 587 & 734 & \(r_e=1.9293\) & \(0, 1\) & \cite{cccbdb} &  \\
\(\mathrm{TiCl}\) & 256 & 320 & \(r_e=2.2697\) & \(3/2, 5/2\) & \cite{aoto2017} &  \\
\(\mathrm{TiH}\) & 256 & 320 & Approximate isotope-derived \(r_e\simeq1.777\) & \(3/2, 5/2\) & \cite{aoto2017} \\
\(\mathrm{TiN}\) & 256 & 320 & \(r_e=1.5802\) & \(1/2, 3/2\) & \cite{aoto2017} &  \\
\(\mathrm{TiS}\) & 256 & 320 & \(r_e=2.0827\) & \(1, 2\) & \cite{aoto2017} &  \\
\(\mathrm{V}\) & 1 & 1 & Isolated atom at origin & \(3/2\) & \cite{ptcomp} &  \\
\(\mathrm{VCl}\) & 256 & 320 & \(r_e=2.2145\) & \(2, 3\) & \cite{aoto2017} &  \\
\(\mathrm{VN}\) & 256 & 320 & \(r_e=1.5703\); Pekeris/Morse inference & \(1, 2\) & \cite{aoto2017} & Vibrational inputs from supporting information \\
\(\mathrm{VO}\) & 256 & 320 & \(r_e=1.5893\) & \(3/2, 5/2\) & \cite{aoto2017} &  \\
\(\mathrm{Zn}\) & 1 & 1 & Isolated atom at origin & \(0\) & \cite{ptcomp} &  \\
\(\mathrm{ZnCl}\) & 256 & 320 & \(r_e=2.13\) & \(1/2, 3/2\) & \cite{aoto2017} &  \\
\(\mathrm{ZnH}\) & 256 & 320 & \(r_e=1.5935\) & \(1/2, 3/2\) & \cite{aoto2017} &  \\
\(\mathrm{ZnS}\) & 256 & 320 & \(r_e=2.0464\) & \(0, 1\) & \cite{aoto2017} &  \\
\midrule
\textbf{Total} & \textbf{25,000} & \textbf{30,270} & 137 compositions & & & \\
\end{longtable}
\endgroup
\noindent The 5,073 source-derived configurations include 46 identity-corrected LAC configurations and two reconstructed primary spectroscopic
parents. A further 19,927 configurations are generated distortions.
The symbols $r_e$, $r_0$, $r_a$, and $r_g$ distinguish equilibrium, effective
ground-state, and diffraction-derived distance conventions.
Additional spin sectors are prescribed vertical-spin training targets at unchanged
nuclear coordinates.
For 46 configurations derived from the LAC vinyl-fluoride parent, we corrected an atomic-number assignment from nitrogen to fluorine, consistent with the parent’s reference geometry.

\section{Appendix 5: Evaluation Targets}
\label{app:evaluation-targets}
$n_{\mathrm a}$ specifies the number of active electrons in the spin-coupling construction. The remaining explicitly treated electrons are assigned to paired spatial orbitals.

\begin{table}[htbp]
\centering
\small
\setlength{\tabcolsep}{6pt}
\renewcommand{\arraystretch}{1.15}
\caption{Target geometries and active-electron counts for the
20,000-update energy comparisons.}
\label{tab:evaluation-targets}
\begin{tabular}{@{}llcc@{}}
\toprule
\textbf{System} & \textbf{Spin state(s)}
& \textbf{Fixed bond length} & $\boldsymbol{n_{\mathrm a}}$ \\
\midrule
H$_2$  & Singlet          & 1.4 bohr~\cite{pachucki2010h2}                    & 0 \\
C      & Singlet, triplet & ---                         & 2 \\
N      & Doublet, quartet & ---                         & 3 \\
O$_2$  & Singlet, triplet & 1.2~\AA                     & 2 \\
O      & Singlet, triplet & ---                         & 2 \\
Sc     & Doublet         & ---                         & 1 \\
Sc$^+$ & Triplet         & ---                         & 2 \\
Ti     & Triplet         & ---                         & 2 \\
Ti$^+$ & Quartet         & ---                         & 3 \\
Cr     & Quintet, septet & ---                         & 6 \\
Cr$^+$ & Sextet          & ---                         & 5 \\
CrH    & Sextet          & 1.6557~\AA~\cite{ref17}      & 7 \\
Fe$^+$ & Quartet         & ---                         & 7 \\
Fe$^+$ & Sextet          & ---                         & 5 \\
Cu     & Doublet         & ---                         & 1 \\
CuH    & Singlet         & 1.4626~\AA~\cite{cccbdb}     & 2 \\
\bottomrule
\end{tabular}
\end{table}

For the 20,000-update energy comparisons, both O$_2$ spin states use the same distance of 1.2~\AA, whereas the spectroscopic reference is adiabatic~\cite{oxygen_nist}.

The calculations in Table~\ref{tab:spin-control} were randomly initialized and optimized for 10,000 updates. Both methods used O$_2$ bond lengths of 1.2156~\AA\ for the
singlet and 1.20752~\AA\ for the triplet. SATE used two active electrons for both O$_2$ states and seven for both Fe$^+$ states. 
\end{document}